\documentclass[twoside,fleqn]{article}
\usepackage{espcrc2}
\usepackage[dvips]{epsfig}
\newcommand\eqa{\begin{equation}}
\newcommand\eqb{\end{equation}}
\newcommand\Eqa{\begin{eqnarray}}
\newcommand\Eqb{\end{eqnarray}}
\newcommand{\subs}[1]{\mbox{\scriptsize\it #1}}   
 
\newcommand{\eqref}[1]{Eq.~(\ref{#1})}            

\newcommand{\fig}[1]{Figure~\ref{#1}}             
 
\newcommand{\mod}[1]{(\mbox{mod}\ #1)}            

\newcommand\Dc{{\cal D}}                          
 
\newcommand\Nb{\bar{N}}                           
\newcommand\rhob{\bar{\rho}}                      
 
\newcommand\rhot{\tilde{\rho}}                    

\newcommand\Nd{\dot{N}}                           

\newcommand\muh{\hat{\mu}}                        

\newcommand\red{_{\protect\subs{red}}} 
\newcommand\eff{_{\protect\subs{eff}}} 
\newcommand\cl{_{\protect\subs{cl}}}
\newcommand\tp{_{\protect\subs{top}}}
\newcommand\CS{_{\protect\subs{CS}}}
\newcommand\Sph{_{\protect\subs{Sph}}}

\newcommand\mmax{_{\protect\subs{max}}}
\newcommand\intS{\int_{0}^{\beta}}

\newcommand\Dx{\Dc[x(t)]}

\title{
{\vspace{-4cm} \normalsize \hfill
\parbox{30mm}{MS-TPI-96-12\\hep-lat/9607052}}\\[30mm]
Topological transitions at $T>0$ in the euclidean 2d U(1)-Higgs model}
 
\author{H. Dilger%
        \thanks{Supported by Deutsche Forschungs-Gemeinschaft} \smallskip\\
        Institut f\"ur Theoretische Physik I,
        WWU M\"unster, Wilhelm-Klemm-Str.\ 9,
        D-48149 M\"unster, Germany}%
 
\begin{document}

\begin{abstract}
The two-dimensional U(1)-gauged Higgs model is studied on an euclidean lattice 
of size $L_1\times L_2$, where the temperature $T=L_2^{-1}$ is of the order of 
the sphaleron mass. The simulation parameters are taken from zero temperature 
results \cite{Jochen}. By comparison with classical and semiclassical 
results I discuss, whether the sphaleron transition rate can be extracted from 
the behavior of the Chern-Simons number and from the formation of vortices in 
an euclidean simulation at high temperatures.
\end{abstract}
 
\maketitle

\section{Motivation}

In the 2d U(1)-Higgs model, as well as in the Standard Model, the transition
between gauge equivalent vacua with integer values of the Chern-Simons number 
$N\CS$ is related to the anomalous violation of the axial U(1)-symmetry. 
The transition rate $\Gamma$ is usually evaluated by semiclassical methods 
\cite{Sphal} and classical real-time simulations \cite{Realtime}. 
It is desirable that an evaluation of $\Gamma$ by euclidean simulations 
gives control over the full quantum corrections to these calculations.

At low temperatures (inverse temperature $\beta\rightarrow\infty$) $\Gamma$ is 
given by the topological susceptibility
\eqa
    \chi\tp = \frac{1}{\beta V} < [N\CS(\beta)-N\CS(0)]^2 > 
             \ \rightarrow \ \frac{\Gamma}{V} \, ,
\eqb
$V$ is the spatial volume. This relation makes use of the random walk 
of $N\CS$ for large $t$.

At high temperatures $\beta$ is too small to see this long-time behavior.
In fact, since the topological charge $N\CS(\beta)-N\CS(0)$ must be an integer,
a configuration contributing to $\chi\tp$ is forced to change $N\CS$
by at least 1 in the short time interval given by $\beta$. This leads to an 
exponential suppression $\chi\tp \sim e^{-c/\beta}$, whereas $\Gamma$ should 
be enhanced by high temperatures \cite{Arnold}.
Can other observables do better?


\section{The high temperature transition rate of the quantum pendulum}

I shall address this question at first for the quantum pendulum, a particle of 
unit mass in the potential $V(x)=[1+\cos(x)]$.
An observable asking for paths from vacuum to vacuum will be suppressed
for $\beta\rightarrow0$, since $\Delta x=2\pi$ in a time $\Delta t=\beta$ is
required.
I thus consider paths, which only need to cross one of the barrier tops 
$V(x)=V\mmax$ at $x_m=(2n\!+\!1)\pi$. I define
\eqa \label{eta-def}
    \rhot \equiv 
    \frac{1}{\beta Z} \int\Dx \ \chi[x(t)] \ e^{-S[x(t)]} \ ,
\eqb
$$
\chi[x(t)] = \left\{ \begin{array}{ll}
     1 & \mbox{if } \exists \ t\in[0,\beta] \mbox{ with } x(t) = x_m \\
     0 & \mbox{else}
                           \end{array} \right. \!\!. 
$$
For any periodic potential I find in the limit $\beta\!\rightarrow\!0$
\eqa
    \rhot \ = \ (\pi/2) \ \Gamma\cl \ ,  
\eqb
the classical transition rate $\Gamma\cl$ is \cite{Affleck}
\eqa
    \Gamma\cl = <\!|p|\, \delta(x\!-\!x_m)\! > 
              = \sqrt{\frac{2}{\pi\beta}} 
              \ \frac{e^{-\beta V\mmax}}{\int_0^{2\pi}\!dx e^{-\beta V(x)}} .
\eqb 
$\rhot$ counts repeated fluctuations over the barrier only once. 
Otherwise it would not be well-defined due to small time fluctuations of 
any quantum path.
However, for small $\beta$ one expects at most one physical transition to 
occur, thus $\rhot$ has the meaning of a transition density per unit euclidean 
time. 
For large $\beta$ and periodic boundary conditions $x(0)\!=\!x(\beta)\!+\!
2\pi n$ the non-transition probability $p\!=\!1\!-\!\rhot\beta$ decays 
as $p\!=\!e^{-\beta\rho}$. So, for the interpretation as a transition 
density I consider 
\eqa
   \rho \equiv -\ln(1-\rhot\beta)/\beta \ .
\eqb 

\begin{figure}[ht]
\vspace{7mm}
\unitlength=0.70mm
\begin{picture}(0,0)
\put(100,-64){\makebox(0,0)[cc]{$\beta$}}
\end{picture}
\epsfysize=52mm
\flushleft{\epsfbox[44 50 680 420]{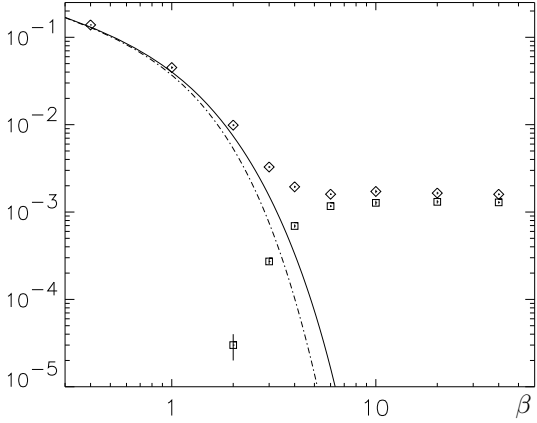}}
\vspace{-26mm}
\caption{$2\rho/\pi$ and $\Gamma$ for the quantum pendulum.}
\label{pd_rho}
\end{figure}

For large $\beta$ the behavior of $\rho$ can be compared with the $T=0$ 
transition rate given by the analogue of the topological susceptibility 
\eqa
    \Gamma_\infty \ = \ \frac{1}{4\pi^2\beta} < [x(\beta) - x(0)]^2 > \ .
\eqb
In \fig{pd_rho} the diamonds give the value of $2\rho/\pi$, the squares 
give the values of $\Gamma_\infty$, the dashed line shows the classical rate
$\Gamma\cl$, the full line shows the rate $\Gamma\Sph$ in sphaleron 
approximation \cite{Affleck}. 

\vspace{-3mm}

There is a qualitative agreement of $2\rho/\pi$ and $\Gamma_\infty$ at
$\beta\rightarrow\infty$. For higher temperatures $\Gamma_\infty$ becomes 
suppressed as expected, whereas $2\rho/\pi$ approaches the classical behavior 
of $\Gamma$. However, I yet didn't manage to improve $\rho$ such that there is 
an exact correspondence to $\Gamma$ on all temperature scales. This question 
shall be addressed in a forthcoming paper.

\section{Topological transitions in the 2d U(1)-Higgs model}

Can a similar observable be defined in the 2d U(1)-Higgs model? Lattice 
formulation and scaling behavior of this model at $T=0$ are described 
in \cite{Jochen}. Here the quantities of interest are the link fields
\eqa
   B^\mu_x \ = \ -\omega_x + A^\mu_x + \omega_{x+\muh} \ \in [-\pi,\pi] \ ,
\eqb
$A^\mu_x$ is the gauge field, $\omega_x$ is the phase of the scalar field.
The difficulty is the analogue of the sector boundaries.
A naive guess is the requirement $N\CS=n+1/2$ with the lattice definition 
\eqa
    N\CS(0) = \sum_{x_1} \frac{A^1_x}{2\pi}, \ \
    \partial_t^+ N\CS(t) = \sum_{x_1} \frac{F_x}{2\pi} \ ,
\eqb
$F_x/(2\pi)$ is the topological density, see \cite{Jochen}.
In addition a topological transition shows a vortex
\eqa
    B^1_x + B^2_{x+e_1} - B^2_x - B^1_{x+e_2} \ \not\in [-\pi,\pi] \ .
\eqb 

However, the above condition $N\CS=n+1/2$ is in general not related to the 
formation of a vortex. Consider the related quantity 
\eqa
    N_B(t) \equiv \frac{1}{2\pi} \sum_{x_1} B^1_{(x_1,t)} = N\CS(t)\ \mod{1}\ .
\eqb
In the continuum model with fixed scalar field length $\rho(x)=v$, $N_B(t)$ 
decouples from all other degrees of freedom. Its effective action reads
\eqa
    S\eff[N_B] \ = \ \frac{2\pi^2}{e^2 L_1} \, \int dt \, 
                     \left( \Nd_B^2 \ + \ v^2e^2 \, N_B^2 \right) .
\eqb
This leads to a probability distribution of the constant mode 
$\Nb\CS = \intS dt N\CS(t) / \beta$
\eqa
    P(\Nb\CS) \ \sim \ \sum_k e^{-\frac{d\beta}{L_1}(\Nb\CS-k)^2}, 
            \ \ d = 2\pi^2 v^2 \, .
\eqb
With growing $L_1$ it becomes constant. Thus, even with fixed $\rho(x)$, i.e.\ 
with parameters which do not allow for the standard instanton or sphaleron 
solutions, configurations with $N\CS\simeq n+1/2$ are not suppressed.
I found this $L_1$-dependence of $P(\Nb\CS)$ also in lattice simulations with 
variable scalar field length, completely dominating the effects induced by 
vortices. Only the parameter $d$ had to be matched.
So the crossing of $N\CS(t)$ through the points $N\CS=n+1/2$ is no good
condition for topological transitions, see also \cite{Karsch}.

Another possibility is to consider the density of vortices per unit time and 
spatial volume
\eqa
    \rho_V \ = \ < N_{\subs{vortices}} > \ / \ (\beta L_1) \ .
\eqb
Again it is important not to count nearby vortices and antivortices separately,
which tend to occur in small clusters. I evaluated $\rho_V(\beta)$ in the 
`Higgs region' of parameter space with the MC algorithm described in
\cite{Jochen}. The $T=0$ results $am_H=0.442(19)$, $am_V=0.258(6)$,
$v=1.849(1)$ lead to a sphaleron energy $aE\Sph=1.01(4)$.

\begin{figure}[ht]
\vspace{11mm}
\unitlength=0.70mm
\begin{picture}(0,0)
\put(98,-62){\makebox(0,0)[cc]{$\beta$}}
\end{picture}
\epsfysize=50mm
\flushleft{\epsfbox[44 50 680 420]{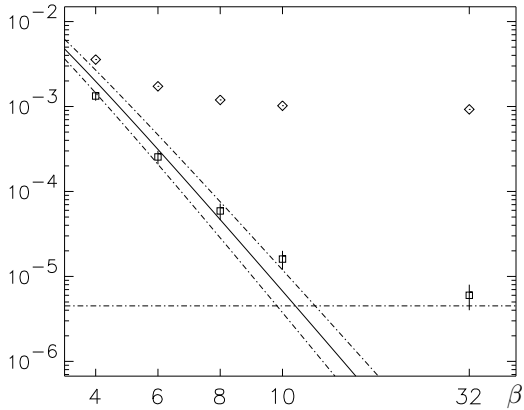}}
\vspace{-26mm}
\caption{$\rho_V$, $\rho\red$, and sphaleron rate $\Gamma_{Sph}/L_1$.}
\label{u1h_red}
\vspace{-6mm}
\end{figure}

\begin{figure}[b]
\begin{flushleft}
\vspace{-6mm}
\unitlength=0.50mm
\linethickness{0.4pt}
\begin{picture}(70,70)(-20,0)
\put( 0, 5){\line(1,0){70}}
\put( 5, 0){\line(0,1){70}}
\put( 0,35){\line(1,0){70}}
\put(35, 0){\line(0,1){70}}
\put( 0,65){\line(1,0){70}}
\put(65, 0){\line(0,1){70}}
\put( 5, 5){\circle*{2}}
\put( 5,35){\circle*{2}}
\put( 5,65){\circle*{2}}
\put(35, 5){\circle*{2}}
\put(35,35){\circle*{2}}
\put(35,65){\circle*{2}}
\put(65, 5){\circle*{2}}
\put(65,35){\circle*{2}}
\put(65,65){\circle*{2}}
\put(20, 2){\makebox(0,0)[cc]{$0$}}
\put(50, 2){\makebox(0,0)[cc]{$0$}}
\put(20,32){\makebox(0,0)[cc]{$\pi$}}
\put(50,32){\makebox(0,0)[cc]{$\pi$}}
\put(20,68){\makebox(0,0)[cc]{$0$}}
\put(50,68){\makebox(0,0)[cc]{$0$}}
\put( 2,20){\makebox(0,0)[cc]{$0$}}
\put( 2,50){\makebox(0,0)[cc]{$0$}}
\put(32,20){\makebox(0,0)[cc]{$\pi$}}
\put(32,50){\makebox(0,0)[cc]{$\pi$}}
\put(68,20){\makebox(0,0)[cc]{$0$}}
\put(68,50){\makebox(0,0)[cc]{$0$}}
\put(35,35){\line(1,1){8}}
\put(42,47){\makebox(0,0)[ll]{$\rho_x\!=\!0$}}
\put(20,50){\circle{12}}
\put(19.5,55.7){\vector(-1,0){1}}
\put(20,50){\makebox(0,0)[cc]{$+1$}}
\put(50,20){\circle{12}}
\put(50.5,25.7){\vector(1,0){1}}
\put(50,20){\makebox(0,0)[cc]{$-1$}}
\end{picture}
\vspace{-11mm}
\end{flushleft}
\caption{$B^\mu_x$ for a simple DVP. Except for the middle site $\rho_x=\rhob$.
} 
\label{lvap}
\vspace{1mm}
\end{figure}
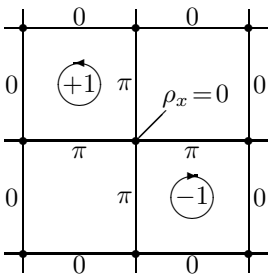

A comparison of $\rho_V(\beta)$ (diamonds) with the transition rate per volume 
$\Gamma\Sph/L_1$ in sphaleron approximation \cite{Sphal} (full line with dashed
error) shows a plateau of $\rho_V(\beta)$ for large $\beta$, see \fig{u1h_red}.
It exceeds the topological susceptibility at large $\beta$ (horizontal line) 
by far. This plateau is related to dislocation-like vortex pairs (DVPs). 
These objects have a size of a few lattice spacings, but fixed $\delta S$ 
for $a\rightarrow0$ compared to a vacuum configuration. A typical example is 
shown in \fig{lvap}. 
For this example I find $\delta S=4\kappa\rhob^2$, compared with a vacuum 
configuration scalar field length $\rho_x\!=\!\rhob,\, B^\mu_x\!=\!0$. 
The DVPs contribute to the expectation value $\rho_V$ and therefore destroy 
its scaling behavior in the continuum limit. Thus it is crucial to separate 
the true topological transitions from these effects on the cut-off scale.%

\begin{figure}[htb]
\vspace{13mm}
\unitlength=0.70mm
\begin{picture}(0,0)
\put(40,82){\makebox(0,0)[cc]{$p(\epsilon\mmax)$}}
\end{picture}
\epsfysize=48mm
\centerline{\epsfbox[0 50 680 420]{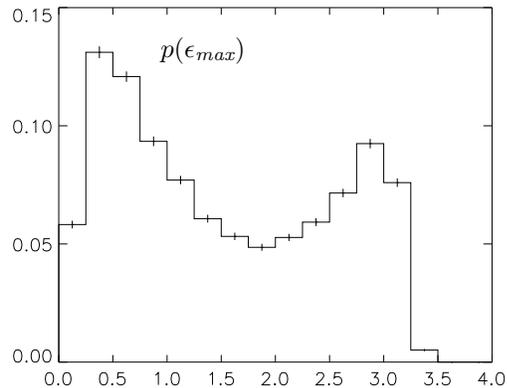}}
\vspace{-25mm}
\caption{$\epsilon\mmax$-histogram for $\beta=4$.}
\label{hist4}
\end{figure}

Under cooling dislocation-like objects should loose their energy density 
faster than physical objects of the size of the correlation length. 
In fact, there is a correlation between vortices and lumps in the energy 
density $\epsilon(x)$ after some cooling sweeps.
\fig{hist4} shows the distribution $p(\epsilon\mmax)$ of the local maxima of 
the energy density near a vortex.
I define the minimal required energy density for a true topological transition 
in the valley between the two peaks, thus throwing away a part of the vortices,
identified as the dislocation-like ones.
The cooling parameters are fixed for all $\beta$-values, a detailed study of
the behavior under cooling shall be given in a future publication.

\vspace{-4mm}

The such reduced vortex density $\rho\red$ (squares in \fig{u1h_red}) fits 
better to the sphaleron rate at small $\beta$ and to the topological 
susceptibility at large $\beta$.
However, this is a rough estimate, far from giving quantitative results. 
Rather it gives a hint which quantities should be better understood even in 
this simple toy model for the high temperature physics of the Standard Model.

\end{document}